\documentclass{aastex}
\usepackage{emulateapj5}
\slugcomment{Revised for the Astrophysical Journal, July 11, 2011}

\shorttitle{Transits of HAT-P-11b}
\shortauthors{Deming et al.}

\begin{document}


\title{Kepler and Ground-Based Transits of the Exo-Neptune HAT-P-11b}


\author{Drake~Deming\altaffilmark{1,2}, Pedro~V.~Sada\altaffilmark{3,4},
Brian~Jackson\altaffilmark{1,4,5}, Steven W. Peterson\altaffilmark{6},
Eric Agol\altaffilmark{7,8,9}, Heather A. Knutson\altaffilmark{8,10},
Donald~E.~Jennings\altaffilmark{1,4}, Flynn~Haase\altaffilmark{6}, \&
Kevin Bays\altaffilmark{6}}


\altaffiltext{1}{Planetary Systems Laboratory, NASA's Goddard Space Flight Center, 
  Greenbelt MD 20771}
\altaffiltext{2}{Present address: Department of Astronomy, University of Maryland at 
  College Park, College Park, MD 20742; ddeming@astro.umd.edu}
\altaffiltext{3}{Universidad de Monterrey, Monterrey, M\'exico}
\altaffiltext{4}{Visiting Astronomer, Kitt Peak National Observatory, National Optical Astronomy 
  Observatory, which is operated by the Association of Universities for Research in Astronomy under 
  cooperative agreement with the National Science Foundation}
\altaffiltext{5}{NASA Postdoctoral Fellow}
\altaffiltext{6}{Kitt Peak National Observatory, National Optical Astronomy Observatory, 
   which is operated by the Association of Universities for Research in Astronomy under 
  cooperative agreement with the National Science Foundation}
\altaffiltext{7} {Department of Atronomy, University of Washington, Box 351580, Seattle, WA 98195}
\altaffiltext{8} {Department of Astronomy, University of California at Berkeley, Berkeley, CA 94720-3411}
\altaffiltext{9} {Miller Visiting Professor at UC Berkeley}
\altaffiltext{10} {Miller Fellow}


\begin{abstract}
We analyze 26 archival Kepler transits of the exo-Neptune HAT-P-11b,
supplemented by ground-based transits observed in the blue (B-band)
and near-IR (J-band).  Both the planet and host star are smaller than
previously believed; our analysis yields
$R_p=4.31R_{\oplus}\pm0.06R_{\oplus}$, and $R_s = 0.683R_{\odot}\pm
0.009 R_{\odot}$, both about $3\sigma$ smaller than the discovery
values.  Our ground-based transit data at wavelengths bracketing the
Kepler bandpass serve to check the wavelength dependence of stellar
limb darkening, and the J-band transit provides a precise and
independent constraint on the transit duration.  Both the limb
darkening and transit duration from our ground-based data are
consistent with the new Kepler values for the system parameters. Our
smaller radius for the planet implies that its gaseous envelope can be
less extensive than previously believed, being very similar to the H-He
envelope of GJ\,436b and Kepler-4b. HAT-P-11 is an active star, and
signatures of star spot crossings are ubiquitous in the Kepler transit
data.  We develop and apply a methodology to correct the planetary
radius for the presence of both crossed and uncrossed star spots.
Star spot crossings are concentrated at phases -0.002 and +0.006.
This is consistent with inferences from Rossiter-McLaughlin
measurements that the planet transits nearly perpendicular to the
stellar equator.  We identify the dominant phases of star spot
crossings with active latitudes on the star, and we infer that the
stellar rotational pole is inclined at about $12^{\circ}\pm5^{\circ}$
to the plane of the sky.  We point out that precise transit
measurements over long durations could in principle allow us to
construct a stellar Butterfly diagram, to probe the cyclic evolution
of magnetic activity on this active K-dwarf star.
\end{abstract}


\keywords{stars: planetary systems - transits - techniques: photometric}

\section{Introduction}

The exo-Neptune HAT-P-11b (\citealp{bakos}, hereafter B10) is prominent
among extrasolar planets smaller than Saturn.  HAT-P-11b transits a
bright star (V=9.59) that lies in the Kepler field \citep{borucki}.
Based on its position in a mass-radius diagram (e.g., Figure~14 of
B10), HAT-P-11b is likely to have a massive atmosphere.  Moreover, B10
found good mass and radius agreement with metal-rich models for the
planet \citep{baraffe}, and the B10 spectroscopic analysis of the host
star indicated that it was metal-rich.  The planet's atmosphere is
therefore likely to exhibit a significant molecular absorption
spectrum during transit and/or eclipse.  It is a tempting target for
future spectroscopic characterization, for example using precise
ground-based spectrophotometry (e.g., \citealp{bean}) in combination
with HST \citep{pont09}, and/or Warm Spitzer \citep{desert}. Prior to
such efforts, it is important to improve our current knowledge of the
system parameters and optical planetary radius by examining the Kepler data.

Based on photometric variations of the star, B10 concluded that star
spots were common on the stellar photosphere. Rossiter-McLaughlin
observations of the system \citep{winn, hirano} indicate that the
planet's orbital angular momentum vector is nearly perpendicular to
the orbital angular momentum vector of the star.  \citet{winn} predicted that
this mis-alignment would produce a characteristic signature in the
spot-crossing patterns seen during transit, and this should be quite evident in
the Kepler data.

In this paper, we report an analysis of Q0-Q2 archival Kepler data for
transits of HAT-P-11b, supplemented with new ground-based transit data
at wavelengths bracketing the Kepler bandpass. The potential benefits
of ground-based transit photometry as a complement to Kepler have been
emphasized by \citet{colon}. 

% Also, because the distance of HAT-P-11 is known to good precision from
% Hipparcos, we measure the absolute infrared flux of the system using
% Warm Spitzer, as a check on the stellar radius and effective
% temperature.

\section{Observations and Photometry}

\subsection{Ground-based Observations}

We observed a transit of HAT-P-11b from Kitt Peak on UT date June 1,
2010, at two wavelengths approximately bracketing the Kepler bandpass.
For the longer wavelength, we used the 2.1-meter reflector with the
FLAMINGOS 2048x2048-pixel infrared imager \citep{elston}, and a J-band
(1.25\,$\mu$m) filter, at 0.6 arc-sec per pixel scale. Following the
conclusion of nightly public programs, we have access to the 0.5-meter
telescope at the Kitt Peak Visitor Center (VCT).  Simultaneous with
the 2.1-meter J-band observations, we observed the same transit at
short wavelength using the VCT and a 3072x2048 CCD camera at 0.45
arcsec per pixel. The VCT observations were acquired in an AstroDon
B-band filter (390-510 nm).  Observations at both telescopes used a
defocus to about 10 arc-sec diameter to improve the photometric
precision, and both used off-axis guiding to maintain pointing
stability.  Exposure times were 30-seconds at the 2.1-meter, and
60-seconds at the VCT.  All of the optical CCD exposures at the VCT
were binned 2x2 to facilitate rapid readout.

Flat-field observations were acquired using either twilight sky (VCT),
or a series of night-sky FLAMINGOS exposures including pointing
offsets to allow removal of stars via a median filter.

Our HAT-P-11 transits were observed on the 6th night of a 7-night
contiguous run on the 2-meter telescope.  At the outset of each
observing run, we check the FLAMINGOS instrument clock to be certain
that it is synced with GPS time signals.  Analyzing the transit data
from this May-June 2010 run, we have discovered large ($\sim 300$
second) differences, exceeding the random errors, between transit
times for other planets measured simultaneously at the VCT {\it vs.}
the 2-meter.  These anomalies occur during the last half of the
observing run, and unfortunately the FLAMINGOS instrument clock was
not re-checked at the end of the run.  Although we have no direct
evidence of clock errors, we conservatively regard our observed J-band
or B-band transit times as suspect, and we omit these transit timings
from the updated ephemeris described in Sec.~6. The photometric shapes
of our ground-based transit curves are not affected by this issue.

\subsection{Photometry on the Ground-based Data}

Subsequent to dark current subtraction and division by a flat-field
frame, we performed aperture photometry on the target star and
comparison stars.  The 20-arcmin field of FLAMINGOS provided 6
comparison stars of comparable IR brightness to HAT-P-11. We used a
circular aperture 31-pixels in diameter (18.6 arc-sec diameter, 9.3
arc-sec in radius) to measure the stars. We varied the aperture size
to optimize the precision in the ratio of HAT-P-11 to the ensemble of
comparison stars.  We subtracted sky background using an annulus
having an inner radius of 18 pixels and an outer radius of 25 pixels.
Normalizing HAT-P-11 to the comparison stars yielded a transit light
curve with an observed scatter for the unbinned data that varied from 0.0016 before transit
to 0.0008 after transit, due to the decreasing airmass during the
observations.  The latter precision (0.0008) is comparable to the
current best ground-based photometry in the J-band
(e.g., \citealp{croll}). We estimated an error for each HAT-P-11 J-band
photometric point as the standard deviation of the ratio to the
individual comparison stars, divided by the square root of their
number (error of the mean).  All of our quoted error and precision
values are given as linear ratios to the average intensity value, not
as magnitudes.

Our photometric aperture for HAT-P-11 nominally excludes a faint
companion (2MASS19505049+4805017) lying 10.7 arc-sec North of
HAT-P-11.  However, because of the defocus, some flux from the
companion will fall within the HAT-P-11 aperture.  Fortunately, the
companion is 6.0 magnitudes fainter than HAT-P-11 at J, and 6.2
magnitudes fainter in the optical (Kepler magnitude). There is no
information available as per possible variability of the companion,
but we see no temporal anomalies in our ground-based data (nor in the
Kepler data for HAT-P-11). The principal known effect of the companion
is that the normalizing (out-of-transit) flux for our HAT-P-11
photometry may be overestimated by as much as 1.003.  Fortunately,
even the maximum contamination will have a negligible effect on our
results.  Consider an out of transit flux equal to $1+\epsilon$, where
$\epsilon$ is the contribution of the companion.  Let the in-transit
flux be $1+\epsilon-\delta$, where $\delta$ is the depth of the
uncontaminated transit.  The mis-normalized in-transit flux is
therefore $(1+\epsilon-\delta)/(1+\epsilon) \approx
1-\delta+{\epsilon}{\delta}$.  In the case of HAT-P-11
we have $\delta \approx 0.004$, and possible contamination $\epsilon
\approx 0.003$.  The second-order term will affect no more than
$\pm1$ in the least significant digit of our results, even for our
Kepler analysis (see below).  We therefore ignore possible
contamination from 2MASS19505049+4805017.

After normalizing to the comparison stars, the HAT-P-11 J-band
photometry exhibited low-amplitude (0.0005) parabolic curvature in the
out of transit baseline. It is probable that this baseline curvature
is caused by differences in the effective wavelengths of the
J-bandpass as a function of stellar color, in combination with the
wavelength-variation in telluric water vapor absorption. This is a
familiar effect that we have seen in other J-band photometry
\citep{sada}.  We removed this baseline curvature using a 2nd-order
polynominal fit, masking off the in-transit portion.  Because the
baseline intensity varies slowly with wavelength, it does not affect
the point-to-point scatter in the transit curve.  However, baseline
uncertainty does represent a potential source of systematic error for
the transit depth, as discussed in Sec.~8.

Photometry of the VCT B-band data used similar procedures as for the
FLAMINGOS J-band data, except that only two comparison stars were
usable, but no quadratic baseline correction was needed.  Figure~1
shows our ground-based transit data for both bands, in comparison to
transit curves calculated using the analytic relations of
\citet{mandel}, with the B10 system parameters, and limb-darkening
coefficients from Kurucz model atmospheres (see Sec.~4).

\section{Kepler Observations}

We analyzed public archival data for HAT-P-11 in quarters Q0-Q2.  Our
analysis uses the short cadence \citep{gilliland} Pre-Search Data
Conditioned (PDC) light curves from the MAST archive. We remove
outlying photometric points from the light curves using a $4\sigma$
clip applied to the difference between the PDC light curve and a
5-cadence median of that same light curve. The brightness of HAT-P-11
varies slowly due to the rotation of star spots with a 29-day period
(B10).  In principle, this brightness variation could be exploited to
correct for the presence of those spots \citep{czesla}.  However, the
particular circumstance of the HAT-P-11b transit motivates a better
method to implement a star spot correction (explained below).  We
therefore remove and discard the stellar brightness variations that
bracket each transit.  We isolate a section of light curve spanning
the center of each transit by $\pm$2.3-hours.  Masking off the
transit, we fit a straight line to the stellar variation over that
section, and ratio the transit light curve to that straight line.

We here verify that a straight line adequately represents the stellar
rotational variability over the 4.6-hour duration of each transit
event. As part of our investigation into the noise properties of the
data (see below), we fit straight lines to 4.6-hour sections of the
data centered on arbitrary planetary orbital phases where no transit
or eclipse occurs. Because any residual curvature could vary from
convex to concave, we average the absolute value of the deviations
from each fitted straight line. We fit a parabola to the average of
those absolute deviations.  We repeated this test for 14 different
phases in the planetary orbit, avoiding the transit itself.  In the
worst of the 14 cases, the maximum span of that parabola over
4.6-hours is less than 4 parts-per-million (ppm), and baseline
excursions of that magnitude will have negligible impact on our
analysis. The fact that the planetary orbit is not phased to the
stellar rotation period implies that baseline effects from stellar
rotation will be further reduced when stacking multiple transits.  We
conclude that straight line baselines are adequate for extracting
transits of the planet over 4.6-hour sections of the Kepler
data. Figure~2 shows an example of a transit with the linear baseline
removed, as well as an additional transit illustrated at the stage
prior to removal of the linear baseline.

\subsection{Noise Properties of the Kepler Data}

Our results for HAT-P-11 are heavily dependent on the Kepler data, so
it is prudent to investigate the noise properties of these data,
especially in the limit where they are averaged to achieve very high
photometric precision. By noise properties, we mean not only the low
amplitude artifacts inherent in the data themselves \citep{gilliland},
but also the existence of fluctuations caused by stellar activity,
such as low amplitude flares \citep{walkowicz}.  A conventional method
to evaluate the noise properties of photometric data is to measure the
standard deviation of binned data as a function of the number of
points that are binned.  For stationary white noise, we expect that
the standard deviation of fluctuations about a mean value will
decrease as the inverse square-root of the number of binned
points. Standard deviations greater than this scaling law are
attributed to so-called red noise \citep{pont}. 

To investigate the noise properties of the Kepler data, we performed
both a conventional binning test, as well as a more specialized test.
Both tests begin by extracting twenty six 4.6-hour sections of the
data centered at an arbitrary phase, i.e., {\it not} centered on
transit or secondary eclipse. We fit and remove a straight line to
each of these data sections, and bin the residuals in two ways.
First, we bin consecutive points in each section, compute the standard
deviation of the binned data, and average the standard deviations over
the 26 sections.  Second, we bin the data non-consecutively, by
combining points from different sections into a common phase bin whose
width is approximately one cadence (60-seconds, 0.00015 in phase).
Figure~3 shows the results from both binning procedures.  The standard
deviation for consecutive binning decreases as $N^{-0.34}$, indicating
the presence of red noise in the data.  However, the standard
deviation for non-consecutive binning agrees very precisely with the
$N^{-0.5}$ relation, indicating that noise components do not persist
at a specific phase from one orbital period to another. Our procedure
for combining transits (see below) shifts them to a common phase,
i.e., we use non-consecutive binning.  Although red noise in the data
will increase the scatter for each individual transit, the noise in
the phased and averaged transit decreases as $N^{-0.5}$, where $N$ is
the number of points in each bin of the combined transit.

\subsection{Removal of Transited Star Spots}

Our analysis includes 26 transits of HAT-P-11 from the Kepler data and
{\it all} of them show signs of star spot crossings, albeit less
prominent than the Figure~2 example.  We have corrected these
HAT-P-11b transits for the presence of spots crossed by the planet as
well as spots {\it not} crossed by the planet.  We here describe the
correction for crossed spots; the uncrossed spot correction we defer
to Sec.~7.

Our correction methodology for crossed spots begins by combining the
26 transits, i.e., transforming them to a common time frame.  In so
doing, we must minimize interference by the crossed spots to our
process of time-shifting the transits.  We note that the effect of
star spots will be greatly reduced at ingress and egress by two
favorable factors.  First, spots during ingress and egress will be
foreshortened by the limb-viewing geometry. Second, we expect few
spots to occur at ingress and egress due to the nature of star spot
distributions.  The R-M results \citep{winn,hirano} indicate a large
angle between the stellar equator and the planetary orbital plane.
Hence ingress and egress will probably occur near the poles of the star. We
expect few star spots at the stellar poles, by analogy with our Sun,
but we acknowledge that polar spots do occur on some active stars
\citep{waite}.

To exploit these favorable factors, we use the ingress and egress data
to determine the center of temporal symmetry for each transit by
constructing bisectors of the transit curve at six intensity levels:
0.9995, 0.9990, 0.9985, 0.9980, 0.9975 \& 0.9970.  We average those
bisectors to obtain the temporal center of symmetry for each transit.
We use those temporal centers to compute an updated ephemeris for the
planet (Sec.~6).  The deviations of individual transits from the new
ephemeris is typically about 10-seconds, with no evidence for real
timing variations.  We therefore use the new ephemeris to combine the
transits onto a common time frame.  This combined transit is shown in
Figure~4, with the abscissa expressed as orbital phase relative to
transit center.

Active regions on our Sun are known to concentrate at `active
latitudes' (sometimes called `preferred latitudes') located
symmetrically about the solar equator.  Figure~4 shows that the effect
of HAT-P-11b's star spot crossings cluster at two orbital phases
(-0.002 and 0.006), which we interpret as the planet crossing two
active latitudes.  This view is consistent with the expected
distribution of star spots, as well as the R-M results \citep{winn,
hirano} that indicate the planet is orbiting nearly perpendicular to
the plane of the stellar rotational equator.  Under the assumption
that the active latitudes are located symmetrically about the stellar
rotational equator, we find that the planet crosses the stellar
equator at phase 0.002.  This implies that the angle between the plane
of the sky and the stellar rotational axis is approximately
$12^{\circ}\pm5^{\circ}$ degrees, where we estimate the precision from
the phase-scatter of the two active latitude crossings.

To remove the effect of the active latitude crossings, we exploit gaps
in the longitudinal distribution of the star spots. If spots
completely covered the active latitudes on the star, the signatures of
spot crossings would appear every time the planet crossed an active
latitude.  However, by analogy with our Sun, we expect significant
gaps between spots on an active latitude.  Thus, spot crossings only
occur when the planet crosses an active latitude and the longitude of
the planet and a star spot coincide. Moreover, Figure~4 implies that
active latitudes are broad: the phase (hence, latitude) distribution
of crossed spots in a given hemisphere is not single-valued.  We
therefore proceed to remove the effect of crossed spots as follows.
First, we group the Figure~4 data using two bin sizes.  During the
transit (first to fourth contact), we use a bin width of 0.00015 in
phase (1 minute, about the same as the Kepler short cadence
time). Keeping the bin width short will eliminate significant
distortion of the transit curve shape \citep{kipping}. Out of transit,
we bin the stellar continuum with a bin width of 0.0006 in phase (4.2
minutes).
				   
For each phase bin, we construct a histogram of relative intensitites
within that bin, and we fit a Gaussian to the peak of that histogram.
The spot-corrected intensity for the transit curve at that phase is
taken as the centroid of the Gaussian. To ensure that peaks in the
wings of the histogram do not perturb the Gaussian fit to the main
peak, we do the fitting iteratively.  Following a preliminary Gaussian
fit, we zero those portions of the histogram lying more than
$2.5\sigma$ distant from the Gaussian center, then we re-fit to the
main peak.  The assumption underlying this method is that star spots
are only a perturbation to the dominant intensity at a given phase,
and that the most common intensity observed at that phase is not
affected directly by star spots.  If this assumption were not valid,
then all methods that do not interpolate over the spot-crossing region
from other phases would fail. In that sense our underlying assumption
is no more restrictive than nature herself. Some of the low-lying
points on Figure~4 could be due to bright active-region morphology,
such as plage. Our methodology automatically corrects for plage
perturbations as well as spots.  Figure~5 shows an example of our
histogram prior to zero-ing the wings, with the Gaussian fit to the
peak.

Figure~6 shows the result of binning and spot-correcting the stacked
transits from Figure~3.  The effects of transited star spots are
largely eliminated in the Figure~6 transit curve, except for a group
of noisy points near phase $-0.0025$, and a single outlying point near
phase $+0.0055$. The spot correction is imperfect in the sense that
the scatter over much of the in-transit portion is about twice that
expected from Kepler's photometric precision. Nevertheless, this
combined transit is much less affected by spot crossings than the
individual transit curves. We note that our correction method is fundamentally
statistical in nature, and will produce even better results as
additional Kepler transits of this system become available.

\section{Limb Darkening}

Prior to fitting the ground-based and Kepler data to derive improved
system parameters, it is necessary to consider the effect of stellar
limb darkening.  In fitting to Kepler data for TrES-2,
\citet{kipping_bakos} pointed out that adopting limb-darkening
coefficients from model atmospheres can potentially make the transit
results model-dependent to an undesirable degree.  We have therefore
used both approaches, adopting limb-darkening coefficients from a Kurucz model
atmosphere for HAT-P-11, as well as fitting for them.

B10 inferred stellar parameters for HAT-P-11 of 4500/4.5/0.3 in
Teff/$\log g$/[Fe/H]. We used a Kurucz model atmosphere with
parameters 4500/4.5/0.5\footnote{http://kurucz.harvard.edu/grids.html/gridP05}, because the
Kurucz grid is tabulated at 0.5-increments in metallicity.  We
verified, by comparison with a solar metallicity model, that the
difference between metallicity +0.3 and +0.5 will be negligible for
limb-darkening.  At each of 17 disk positions ($\mu$-values), we
weight the model atmosphere intensities by the filter or Kepler
bandpass functions, and integrate over wavelength.  This defined the
limb darkening for each bandpass.

We fit linear and quadratic coefficients for the standard expressions
to the results of our model atmosphere integrations
\citep{claret}. \citet{sing} has calculated limb darkening
coefficients for the Kepler bandpass, and we compared our results to
his tabulation.  For 4500/4.5/0.5 we find (linear, quadratic)
coefficients of (0.6136, 0.1062) {\it vs.} (0.6266, 0.1057) for
\citet{sing}.  We performed similar calculations for the J- and B-band
filter functions.  Comparing our quadratic coefficients to
\citet{claret} for the J-band filter and 4500/4.5/0.5, we find
reasonable agreement: (0.290, 0.244) for us, and (0.267, 0.255) from
\citet{claret}.  For the B-band filter, we find that the Kurucz limb
darkening can be well approximated using linear limb darkening, i.e.,
with the quadratic coefficient set to zero.

We conclude that our calculations of limb-darkening coefficients
accurately reflect the output of Kurucz model atmospheres.  But
whether the actual star conforms to those calculations must be
determined by comparison to our transit data.

\section{Transit-Fitting Methodology}

\subsection{Priors, and Transit Curves}

The orbit of HAT-P-11 has non-zero eccentricity ($e=0.198$, B10).  In
principle, precise transit light curves may reveal some direct
signature of a non-zero eccentricity such as acceleration between
ingress and egress.  However, detection of the secondary eclipse
(e.g., using Spitzer photometry) is potentially the most sensitive
constraint on the eccentricity of the orbit \citep{charb, deming}.
Spitzer observations of HAT-P-11 spanning the time of secondary
eclipse were obtained by R. Barry (Spitzer program 60063), but those
data are still under analysis.  Indeed, it is possible that the
secondary eclipse of this relatively cool planet will prove too weak
to be detectable in the Spitzer data.  Therefore, for all of our
transit fits, we fix the orbital eccentricity and argument of
periastron at the values derived by B10, except that we alter $\omega$
by 180-degrees, as per the difference between $\omega$ for radial
velocity observations (orbit of the star) and $\omega$ for transits
(orbit of the planet). 

We compute transit curves using a new version of the \citet{mandel}
algorithms.  This new version includes the effects of a non-circular
orbit, calculating the sky-projected distance of the planet from the
center of the stellar disk by solving the elliptical geometry. We also
specifically verified that our code can reproduce the obseved radial velocities for
this system (B10 \& \citealp{hirano, winn}).  The new code is also
faster than previous versions.

\subsection{Markov Chain Monte Carlo Code}

We fit theoretical transit curves to the transit data using a Markov
Chain Monte Carlo (MCMC) algorithm \citep{ford}.  We use the
Metropolis-Hastings algorithm with Gibbs sampling. We adjust the step
size for each variable so as to obtain acceptance rates between 30\%
and 60\%, and we run our chains for $10^6$ samples. Prior to starting
each MCMC chain, we re-scale the error bars for the data to insure
that the best reduced $\chi^2$ will be close to unity. This helps to
insure that the errors calculated from the posterior distributions are
realistic. The re-scaling factor was approximately 2 for the binned
Kepler data.  For the ground-based data, we adopt error bars equal to
the observed scatter in the data, so no re-scaling is necessary.

We discard the first 20\% of each chain when tabulating the posterior
distributions. To verify convergence, we compare the posterior
distributions from four chains that have different starting values and
slightly different step sizes. Since our code is new, we tested it in
several ways.  These tests began with simple numerical problems such
as fitting to an average of a series of numbers, and fitting to data
that scatter around a straight line.  Our final test was to generate
synthetic transit data by adding noise to theoretical transit curves
based on the \citet{mandel} formulae, and fitting to those synthetic
transits to verify that we recover the system parameters that were
used to generate the synthetic data.

\subsection{Kepler Transit Fits}

Ideally, we would like fit to all of the individual transit curves
simultaneously, allowing the central transit times to be free
parameters in the fit.  However, the spot crossings that contaminate
individual transits force us to fit to the binned and cleaned transit
(Figure~6). One caveat to this procedure is that imperfections in the
mutual phasings of the transits could potentially broaden and distort
the binned transit.  To check our result, we combined the transits two
ways.  First, we use our new ephemeris (Sec.~6) to shift each transit
to a common phase.  Second, we use the raw individual central transit
times from the bisector analyses to phase the transits.  We performed
all of our MCMC fits to binned transits constructed using both
methods, and found agreement within the random errors. We are
confident that phasing errors do not contaminate our results to a
significant degree.  We report the fit results from the ephemeris
phasing, because it employs the constraint that the transit times
should be strictly periodic in the absence of planetary perturbations.

Our MCMC fits to the binned and cleaned Kepler data (Figure~6) include
6 variables in the fit: $a/R_s$, $R_p/R_s$, orbital inclination,
quadratic and linear limb darkening coefficients, and a correction to
transit center time.  The latter is expected to be zero because the
stacking and binning procedures aligned the individual Kepler transits
to the common transit-centered time frame.  Within the errors, the
MCMC fits retrieved a central time correction consistent with zero.  We performed
a second independent set of MCMC fits to the Kepler data by fixing the
limb-darkening coefficients at their model-atmosphere values.

We tabulate best-fit values for the system parameters by averaging
over the last 800,000 samples for four independent MCMC chains. As a
check on those best-fit values, we implemented an independent $\chi^2$
minimization solution using a Levenberg-Marquardt algorithm. Our
$\pm1\sigma$ error limits equal the values where high and low-side
tails of the MCMC posterior distributions contained 15.9\% of the total samples. The
error limits were close to symmetric on each side of the distributions
(high- and low-side errors typically agree within 10\%). We
conservatively adopt the greater value as the symmetric error for each
fitted parameter.  The best-fit values and errors are listed in
Table~1, except for the quadratic limb darkening coefficient that has
little impact on our analysis.  The best-fit transit curve is plotted
over the binned transit data in Figure~6, together with the curve
expected from the B10 discovery parameters.

\subsection{J- and B-band Transit Fits}

The ground-based data give us an opportunity to check parameters such
as limb darkening over an extended wavelength range. Moreover, the
greatly reduced limb darkening in the J-band results in sharp ingress
and egress times, and that sharp definition of the transit duration
will prove to be useful, as we discuss in Sec.~8, and as \citet{colon}
predicted.  However, because the ground-based data do not have
photometric precision comparable to the Kepler data, we find it
prudent to restrict the ground-based fits to extract fewer parameters.
For the J-band, we set the quadratic limb-darkening coefficient to
equal the model atmosphere prediction ($0.244$), and we set the B-band
quadratic coefficient to zero as noted above.  Anticipating our
results, we find rough agreement between the fits and the model
atmosphere limb-darkening predictions.  The quadratic portion of the
limb-darkening has a minor effect compared to the linear coefficient,
especially in B-band where the strong limb darkening is well
approximated by the linear law. Note that \citet{southworth} found
linear limb darkening to be adequate for the analysis of high quality
ground-based transit observations. We judge that we lose little
information by our adopted restrictions, but the subsequent
restriction in the fitted parameters helps to increase the usefulness
of the ground-based data.

In the case of the the B-band transit data, we are primarily
interested in the consistency between the retrieved linear
limb-darkening coefficient and the model atmosphere prediction.  So in
that case, we fix both the orbital inclination and the value of
$a/R_s$ to their best-fit Kepler values, and we fit only for the
linear limb-darkening coefficient, as well as $R_p/R_s$ and central transit
time.  As regards the linear limb darkening coefficient, we impose the
restriction that the MCMC chains cannot step to values exceeding
unity, since those values produce unphysical (negative) disk
intensities.  

Our first exploratory MCMC chains for the J-band fit showed a strong
degeneracy between orbital inclination and $a/R_s$.  This is not
surprising, since we have previously highlighted this degeneracy for
small planets \citep{sada}.  In the limit of a small planet transiting
a uniform stellar disk, the transit curve approaches an inverse
square-wave function where the duration of the transit measures only
the total length of the chord across the stellar disk.  In that case,
the impact parameter (i.e., orbital inclination) and stellar radius
can trade-off freely.  Hence, in the J-band we fix the orbital
inclination at the Kepler value ($89.41^{\circ}$, Table~1), and we
solve for $a/R_s$.

Results from the J- and B-band fit procedures are included in Table~1,
and best-fit transit curves are overplotted on Figure~1.

\section{Updated Transit Ephemeris}

A useful by-product of our transit analyses is that we can update the
transit ephemeris for this system. We include transits at the two
epochs reported by B10, as well as transits from \citet{hirano} and
\citet{dittmann}, and the Kepler transits.  Table~2 gives the central
transit times and errors for the Kepler transits, using our bisector
method. The precision of the updated ephemeris is dominated by the
Kepler transits, that each have a timing precision of order 10
seconds.  An error-weighted linear least-squares solution for the
ephemeris yields $T_0=2454605.89155\pm0.00013$, in a barycentric TDB
frame \citep{eastman}, and $P=4.8878018\pm 1.6 \times 10^{-6}$ days.
We consider this to be a provisional update, because many additional
values for transit center times will be possible with future Kepler
data. Figure~7 shows residuals for the times of individual transits,
after removing the best-fit ephemeris.  As noted in Sec.~2.1, we omit
our 2010 J- and B-band transits from the ephemeris solution, but we
include their residuals on Figure~7 where they individually lie off
the best-fit ephemeris, but agree with it on average.

\section{Correction for Uncrossed Star Spots}

In the case where multiple high quality transit light curves are
available for a planet that transits at an oblique angle to the
stellar equator, we can correct the derived planetary radius for the
effect of star spots that are {\it not} crossed by the planet during
transit.  The method we describe here assumes that the distribution of
star spots is not correlated with the transits. We argue that this
method has advantages over inferences based on the rotational light
curve of the star \citep{czesla}, because star spot effects in the
rotational light curve can be reduced when multiple spots are
distributed uniformly over longitude.

The formalism of our star spot correction has two broad steps.  First,
we integrate over the path crossed by the planet, and average over all
observed transits, to calculate the average flux deficit due to star
spots on the path of the planet.  The orbit of HAT-P-11b is
essentially perpendicular (within the errors) to the stellar equator
\citep{winn, hirano}.  We find (Table~1) that the orbital inclination
is very close to 90-degrees, and therefore the impact parameter is
near zero.  Moreover, the stellar active latitudes are not far from
the stellar equator.  Therefore the flux deficit that we calculate in
this first step will be characteristic of regions near the center of
the stellar disk.  The second broad step will extend the flux deficit
calculated over the transit path of the planet, to estimate the total
spot coverage over the entire Earth-facing hemisphere of the star.

Figure~8 shows a cartoon of the transit geometry.  Near disk center,
the planet subtends an approximately constant range of longitude as it
transits (indicated by blue meridians of longitude on Figure~8).  This
approximation of course breaks down near the poles because meridians
of longitude converge, but we expect few if any star spots at the poles.
Hence, by integrating over the path of the planet we are essentially
defining the spot coverage in the range of longitude defined by the
angular extent of the planet.

Consider a highly simplified situation where the planet transits a
very small star spot present on a star without limb darkening. Let
$F_0$ be the flux from the {\it unspotted} star, and let $\delta F$ be
the stellar flux deficit caused by the small star spot. In transit,
before the star spot is crossed, the observed flux $F_{obs}$ is
$F_0-F_0(R_p^2/R_s^2)-\delta F$, where $R_p$ and $R_s$ are the radii of
the planet and star respectively. When the planet crosses the star
spot, the $\delta F$ term vanishes, so the flux becomes
$F_{obs}=F_0(1-R_p^2/R_s^2)$, which is the usual expression for the
in-transit flux of a star neglecting limb darkening.  The effect of
the star spot will be seen in the transit curve as an inverted square wave
of amplitude $\delta F$, lasting for a crossing time (measured in
phase units) $t_{\phi}$.  (Note that we use orbital phase as a time
variable, not an angle.)

A very small star spot would create a simple square-wave type
signature in the transit curve, but real star spots are comparable
to the size of the planet itself, their intensity varies from the
outer penumbra to their umbral core, and they often occur in groups.
Therefore their signature in transit light curves can be complex, not
a simple square wave.  Nevertheless, we can derive the total star spot
flux deficit crossed by the planet during a transit, $\delta F_t$, as:

% \begin{equation}
$$\delta F_t  =  {\int \delta F(\phi) d\phi}/t_{\phi}$$
% \end{equation}

where $\delta F(\phi)$ is the amplitude of the deviation seen in the
transit curve at phase $\phi$, and the integral is taken over the path
of the planet, literally over the transit curve.  We need not
explicitly consider the intensity gradient across a star spot, we can
derive the total flux deficit from the above integral, independent of
star spot morphology.  Moreover, we can evaluate the integral directly
from the observed transit curves; we use the stacked transit curves
(Figure~4). We also calculate that $t_{\phi} = 0.785 D/vP = 0.000997$,
where $v$ is the tangential velocity of the planet in its orbit, $P$
is orbital period, and $D$ is the planet's diameter.  The factor of
$0.785$ allows for the fact that the average chord across a circular
planet is less than the diameter. The numerical integration of the
Figure~4 transits yields $\delta F_t = 0.001037$ in flux units where
$F_{obs} =1$.  This value applies to the flux deficit over a range of
longitude defined by the angular extent of the planet at disk center
(between the two blue meridians on Figure~8).  Over that longitude
range, the total flux deficit due to star spots is (on average), about
0.1\%.  Because $2R_p/R_s \approx 0.118$, the 0.1\% applies to a range
of about 0.118 radians. There are 26 such wedges of longitude on the
entire Earth-facing hemisphere of the star.  Neglecting limb effects,
the total star spot flux deficit could be as large as 2.7\%.

Our second broad step assumes that the average size and abundance of
star spots is independent of disk position, but that the projected
area - hence the flux deficit - of star spots decreases as $\cos
\theta$, where $\theta$ is the angular longitude distance from disk
center.  Our second step will therefore integrate over longitude
($\theta$) to obtain the total flux deficit for the entire
Earth-facing hemisphere of the star.  Thus:

%  \begin{equation}
$$F_0 = F_{obs} + \int^{\pi/2}_{-\pi/2} {\delta F_t {\theta_p}^{-1} \cos(\theta) d\theta} 
= F_{obs} + 2\delta F_t /{\theta}_p$$
%  \end{equation}

where $\theta_p$ is the longitude interval covered by the planet
during transit.  This integration yields $F_0 = 1.0176 F_{obs}$.  So
we calculate that the star spots on HAT-P-11 cause the observed
stellar flux to be, on average, lower by 1.76\% compared to an
unspotted star of the same radius and spectral type.  The peak-to-peak
variations seen in the stellar rotational light curve are about 1.5\%
in the Kepler data. This is somewhat smaller than 1.76\% because the
broad distribution of the spots in longitude reduces their signature
in the rotational light curve.  We note that B10 found a significantly
smaller peak-to-peak rotational light curve amplitude (0.6\%), but
their photometry was obtained 1- to 2-years earlier than our Kepler
data. 

The methodology described above makes approximations that depend on
the fortuitous geometry wherein HAT-P-11b crosses nearly perpendicular
to the stellar equator. Moreover, we also retained the approximation
of neglecting stellar limb darkening.  We believe that a more general
formalism could be developed along the same line of reasoning, that
could be applied to less strongly inclined planets, and could include
realistic limb darkening.  That generalization is beyond the scope of
this paper, and we will utilize our current estimate of the total spot
flux deficit of HAT-P-11 when interpreting our results in the next
Section.

\section{Results and Discussion}

Our MCMC fits produce excellent agreement with the Kepler data
(Figure~6). The retrieved parameters agree closely whether we solve
for limb darkening, or fix the coefficients at their Kurucz model
atmosphere values.  Moreover, the retrieved linear coefficient in the
former case ($u = 0.626 \pm0.014$, Table~1) is in excellent agreement
with the model atmosphere value ($0.6179$).  We conclude that the
HAT-P-11 system parameters derived from the Kepler MCMC fits are
robust, and not model-dependent via limb darkening. \citet{knutson}
reached a similar conclusion in their analysis of HST observations of
HD\,209458b.

The $\chi^2$ value for the best fit solution shown on Figure~6 is 417
for the in-transit points, for 112 degrees of freedom.  Thus, the
error bars based purely on Kepler photometric precision must be
increased by a factor of 2 to account for the imperfect precision of
star spot removal.  That factor was applied in the MCMC fits, as noted
in Sec.~5.2.

Both the J-band and B-band transits imply linear limb darkening
coefficients ($u$ in Table~1) that are in reasonable agreement with
model atmosphere values, but they do hint that limb darkening for the
real star could vary more strongly with wavelength than the model
atmosphere predicts.  The MCMC posterior distribution for $u$ in the
B-band (not illustrated) peaks at unity, and values exceeding unity
are unphysical. The 1-sided Gaussian distribution has $\sigma =
0.080$, so our result (Table~1, $u=1.000 \pm0.080$) falls {\it above}
the model atmosphere value ($0.862$) by $1.8\sigma$.  The J-band
result ($u=0.086\pm0.065$) similarly falls {\it below} the model
atmosphere value ($0.244$) by $2.4\sigma$.  If discrepancies in model atmosphere
predictions for $u$ do vary with wavelength in this fashion (stronger
at short-$\lambda$, weaker at long-$\lambda$), we would not
necessarily expect a significant effect in the Kepler band, because
it lies intermediate in wavelength between our B- and J-band data.
Moreover, any such systematic variation should be confirmed using
transits of larger planets, exhibiting deeper transits, where greater
precision in derived limb darkening can be achieved. We note that
there is observational precedent for limb darkening at short wavelengths to be
stronger than model atmosphere predictions \citep{tingley}. We plan
additional simultaneous B- and J-band observations, of giant planet
transits.

Our results for the J-band transit are inconsistent with the Kepler
results as regards the planetary radius.  In J-band, we find that
$R_p/R_s$ is 6\% larger than the Kepler solution ($0.0627$ {\it vs.}
$0.0589$), and the difference is more than 5 times the precision of
the J-band measurement. We regard the Kepler result as definitive, so
we consider how this discrepancy can be explained.

One potential explanation of a discrepant radius in J-band is that it
reflects a true variation of the planetary radius with wavelength, due
to atmospheric opacity.  However, the difference here seems
implausibly large, and we prefer a more mundane explanation. The
transit of HAT-P-11b is relatively long in duration (2.3 hours), and
shallow (0.004).  Ground-based infrared photometry can be subject to
baseline fluctuations caused by telluric water vapor absorption at the
edges of the JHK bandpasses.  The longer the duration of a transit
event, the more sensitive it is to baseline effects, because the
adopted baseline has to span a longer interval.  Also, a given
baseline error will have a greater relative effect for shallow
transits.  Hence, HAT-P-11 is particularly prone to baseline errors,
and we regard the $R_p/R_s$ value from our J-band MCMC fits as
unreliable at the level of accuracy needed for meaningful comparison
with Kepler results. We note that the discrepancy would be even larger
without the baseline correction that we applied in Sec.~2.2; evidently
our correction underestimates the telluric effects.

Although $R_p/R_s$ from our J-band data is questionable compared to
the Kepler value, we believe that $a/R_s$ is reliable, and a useful
complement to the Kepler results.  As noted in Sec.~5.4, the J-band
transit is sensitive to the total chord length.  Already we have seen
(Figure~1), that the duration of the J-band transit predicted from the
B10 discovery results (blue curve on Figure~1) does not fit the
transit duration seen in the J-band data.  Unlike the situation with
transit depth, the transit duration observed in J-band is insensitive
to the telluric atmosphere.  At this wavelength, the ingress and
egress are sharp and well defined, and rapid changes of this type are
much less likely to be caused by the telluric atmosphere.  Fixing the
orbital inclination at the Kepler value, our J-band fits give $a/R_s
=16.454\pm0.131$, in excellent agreement with the Kepler value of
$16.549\pm0.230$.  Transit determination of $a/R_s$ can be used to
derive the stellar density \citep{seager}, for comparison to
asteroseismology results.  However, the Kepler asteroseismology
results for HAT-P-11 were regarded as preliminary by \citet{cd},
so the comparison is premature.

We adopt our Kepler-1 solution (see Table~1) for the system
parameters, and note that the stellar mass ($0.809M_{\odot}$)is
well-determined from the Hipparcos parallax and the isochrone fits
performed by B10.  This yields $R_s = 0.683R_{\odot}\pm 0.009
R_{\odot}$, a $3.3\sigma$ revision from the B10 value ($0.752
R_{\odot}\pm0.021R_{\odot}$).  Using our $R_p/R_s = 0.05892\pm0.00027$,
yields $R_p=4.39R_{\oplus}\pm0.06R_{\oplus}$, a $2.1\sigma$ revision
from B10 ($R=4.73\pm0.16R_{\oplus}$).  However, the presence of unseen
star spots will cause the planet to appear larger in the transit
solutions by 1.76\%, as discussed in Sec.~7.  We thus correct our radius
for the planet downward to $R_p=4.31R_{\oplus}\pm0.06R_{\oplus}$.
Hence we find that both the planet and star are smaller than the B10
discovery values, at about the $3\sigma$ level.

We have been able to improve the system parameters of HAT-P-11 over
the very thorough analysis by B10, in large part due to the numerous
precise Kepler transits.  These data have allowed us to remove the
effect of star spot crossings, resulting in a deeper transit. Although
the deeper transit would tend to produce a larger planet, we also find
a smaller star and the decrease in stellar size, combined with our
correction for uncrossed star spots, results in a smaller planet
compared to B10.  Also in comparison to B10, we find a longer transit
duration, and a smaller impact parameter since our orbital inclination
is closer to 90 degrees. 

Our revised radius for HAT-P-11b has noteworthy implications for its
internal structure.  A mass-radius diagram for planets similar to
HAT-P-11b is illustrated by \citet{lissauer} (their Fig.~5). The H-He
envelope mass implied by the previous radius of HAT-P-11b is close to
20\%; our new radius reduces the required envelope mass to about 14\%,
very close to the envelopes implied for GJ\,436b and Kepler-4b. 

The radius of HAT-P-11b is now known with sufficient precision to
contemplate useful comparisons with other space-borne transit
observations, such as HST and Spitzer could acquire.  That comparison
could potentially reveal an atmospheric signature of HAT-P-11b.
Moreover, we point out that long-term monitoring of the system by
Kepler, potentially supported by precise ground-based photometry after
Kepler has ended, could reveal fundamental information on the magnetic
cycle of the star.  Because the planet transits nearly perpendicular
to stellar active latitudes, monitoring of the number and orbital
phase of star spot crossings could allow us to construct a stellar
`Butterfly diagram' depicting the cyclic evolution of magnetic
eruptions at the stellar photosphere of this active K-dwarf star.

While this paper was under review, two additional analyses of HAT-P-11b
based on the Kepler data were announced. \citet{southworth11} obtains
system parameters that agree with our Table~1.  He does not explicitly
correct for the effect of star spots, instead treating the spot
perturbations as correlated errors.  An analysis by
\citet{sanchis-ojeda} also finds similar system parameters as per our
Table~1.  These authors correct for crossed spots `by hand', but do
not correct the planetary radius for the effect of uncrossed spots.
Interestingly, \citet{sanchis-ojeda} also discuss a second possible
orientation for the system in which the star is viewed nearly pole-on.
In that orientation, the two preferred phases of star spot crossings
could be caused by one active band of spots encircling the stellar
pole.  While we cannot rigorously exclude that geometry, we note that
our correction for uncrossed spots (Sec.~7) - based on viewing the
star as per the Figure~8 geometry - predicts a peak-to-peak variation
in the stellar rotational light curve of $\leq 1.76\%$, in good
agreement with the variability seen in the Kepler data ($\sim 1.5\%$).
That agreement would have to be accidental if the star is being viewed
pole-on, so we believe that our Figure~8 geometry is correct, and the
stellar pole is nearly in the plane of the sky.

\acknowledgements We thank Dick Joyce for his expert assistance with
FLAMINGOS, and we thank an anonymous referee for insightful comments
that allowed us to improve this paper.  Eric Agol acknowledges support
from NSF Career grant AST-0645416.

\clearpage

%% the file, uncomment the \plotone, \plottwo, and
%% If you need a layout that cannot be achieved with \plotone or
%% \plottwo, you can invoke the graphicx package directly with the
%% \includegraphics command or use \plotfiddle. For more information,
%% please see the tutorial on "Using Electronic Art with AASTeX" in the
%% documentation section at the AASTeX Web site,
%% http://www.journals.uchicago.edu/AAS/AASTeX.
%%
%% The examples below also include sample markup for submission of
%% supplemental electronic materials. As always, be sure to check
%% the instructions to authors for the journal you are submitting to
%% for specific submissions guidelines as they vary from

%% This example uses \plotone to include an EPS file scaled to
%% 80% of its natural size with \epsscale. Its caption
%% has been written to indicate that additional figure parts will be
%% available in the electronic journal.

\begin{figure}
\epsscale{.60}
\plotone{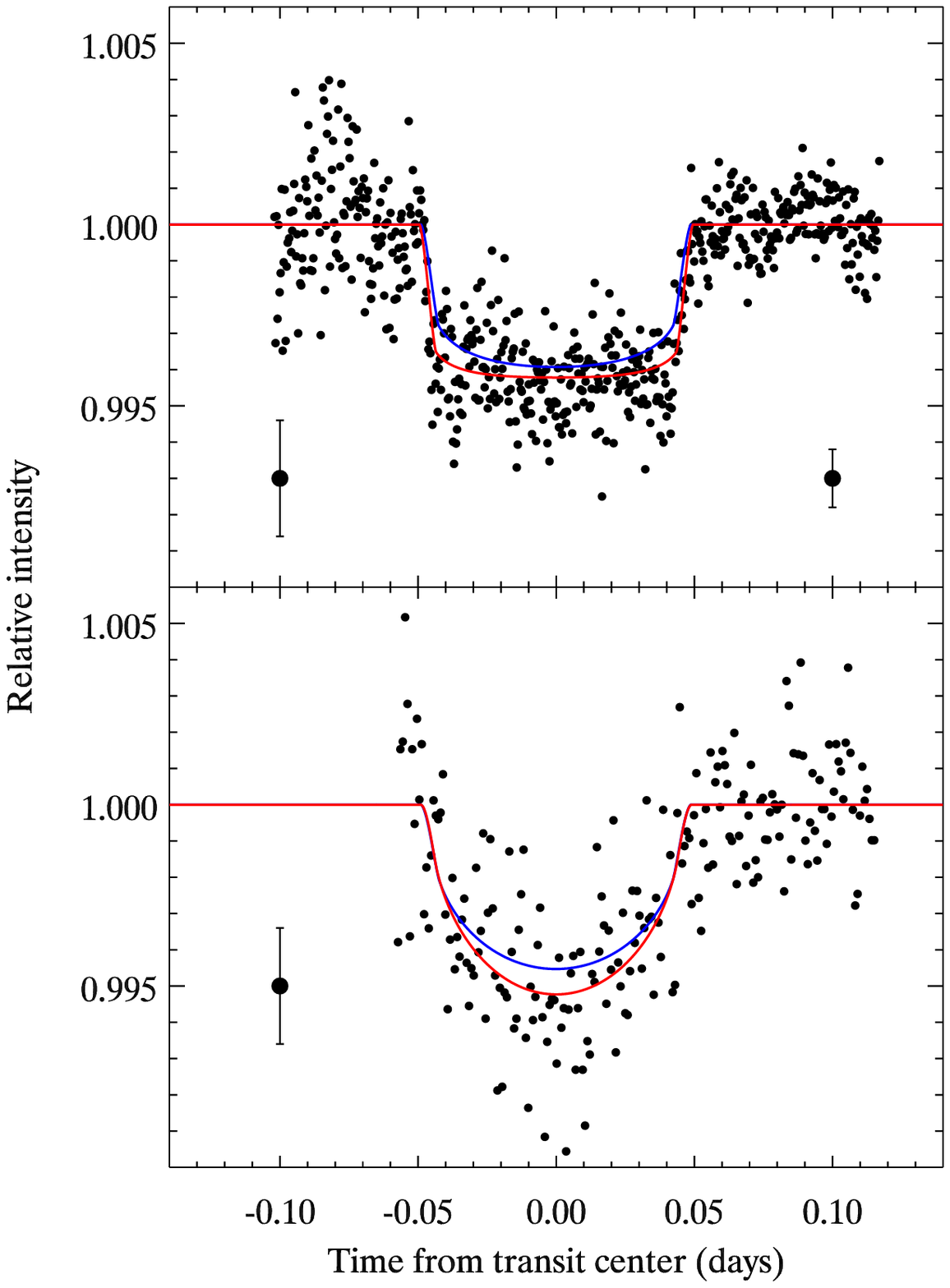}
\vspace{0.7in}
\caption{Transit of HAT-P-11 observed in the J-band (upper panel) and
B-band (lower panel) from Kitt Peak on June 1, 2010.  A quadratic
baseline has been removed from the J-band photometry, as noted in the
text. The error bars per point are indicated by the inset points; for
the J-band observations the error bars decreased after transit due to
decreasing air mass. The overplotted blue curves are the nominal
transit shapes expected using the system paramaters from \citet{bakos}
and limb-darkening from a Kurucz model atmosphere. Note that the
observed duration of transit for the J-band is greater than expected
from the blue curve. The red curves are based on our best-fit
parameters from Table~1.
\label{fig1}}
\end{figure}

\clearpage

\begin{figure}
\epsscale{.50}
\plotone{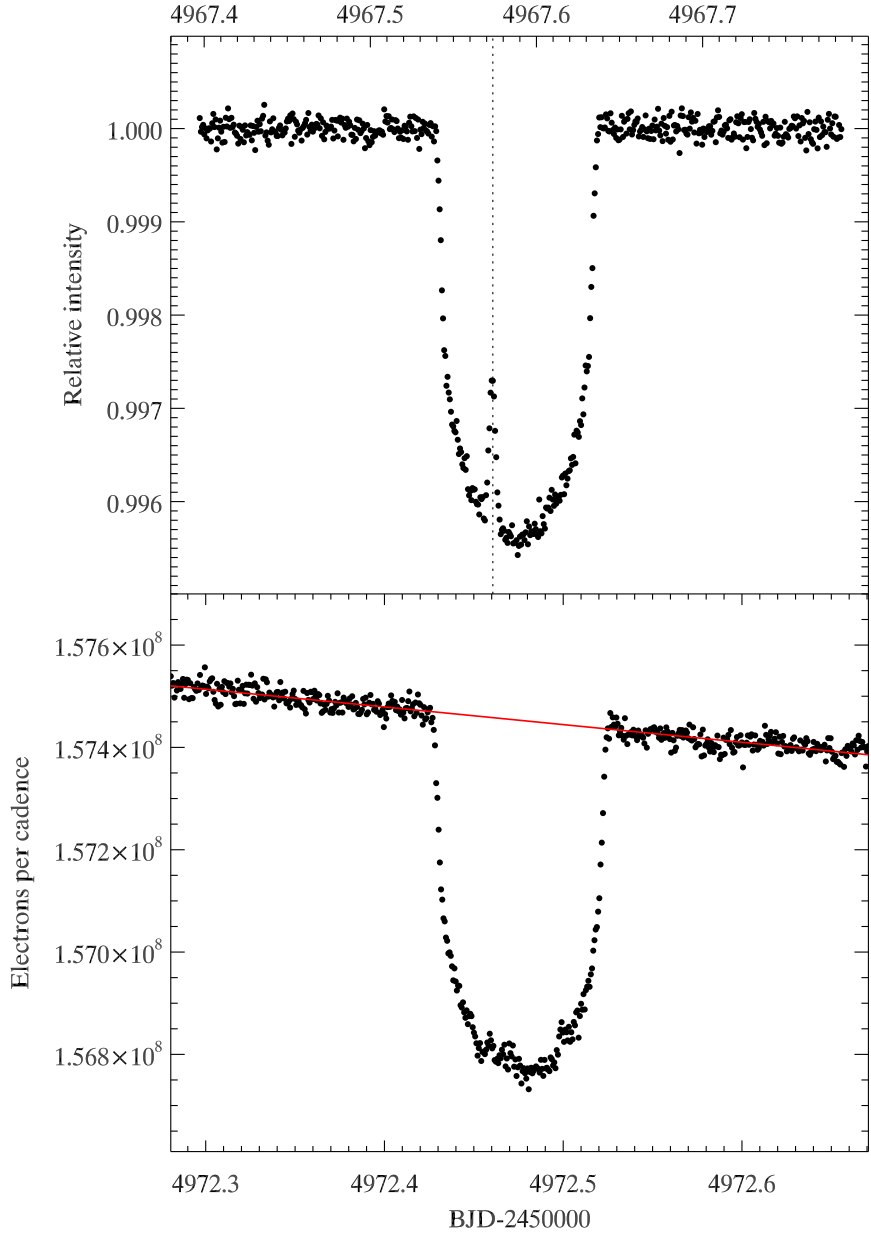}
\vspace{0.6in}
\caption{{\it Upper panel:} Detrended and normalized transit of
HAT-P-11 observed by Kepler on May 15, 2009.  The vertical dashed line
marks a prominent effect due to a star spot crossing. {\it Lower
panel:} the next succesive transit observed by Kepler following the
May 15 transit, shown before normalization and trend removal.  The red
line is the variability of the star, taken to be linear over the span of the transit.
\label{fig2}}
\end{figure}

\clearpage

\begin{figure}
\epsscale{.80}
\plotone{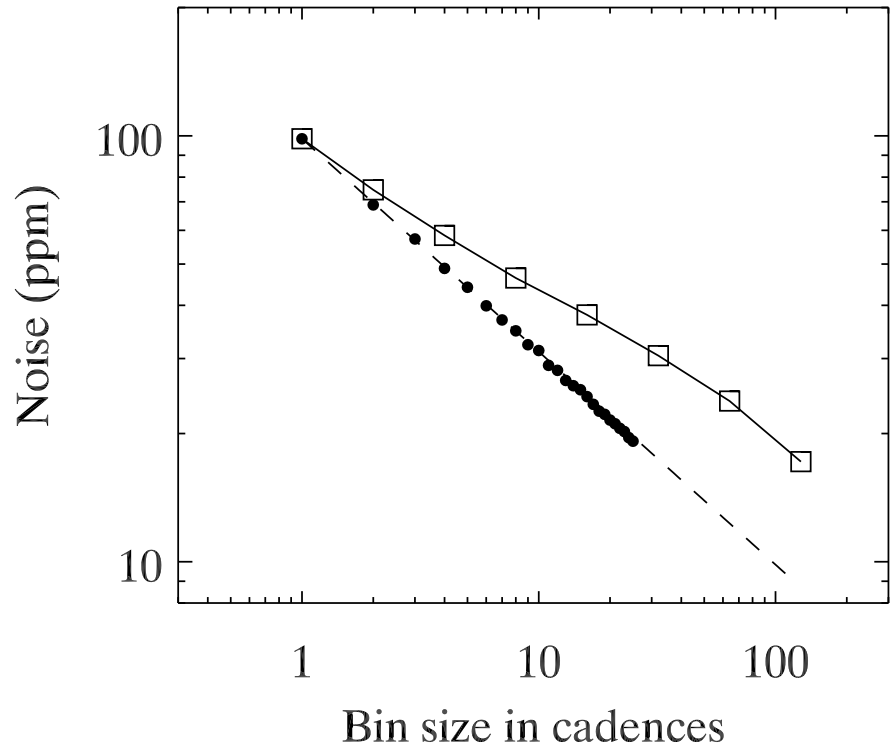}
% \vspace{0.8in}
\caption{Noise properties of the HAT-P-11 short cadence data, showing
standard deviation (ordinate, in parts-per-million) of binned data
versus bin size (abscissa).  The open squares are for binning
consecutive data points, and lie above the theoretical relation
(dashed line) due to correlated red noise of low amplitude. The solid
points show noise amplitudes from bins at a given phase, constructed from $N$ 4.6-hour
sections of data, versus $N$.  These points lie almost exactly on the
theoretical relation (dashed line), showing that noise is uncorrelated
from one orbit of HAT-P-11 to another.
\label{fig3}}
\end{figure}
\clearpage

\begin{figure}
\epsscale{.70}
\plotone{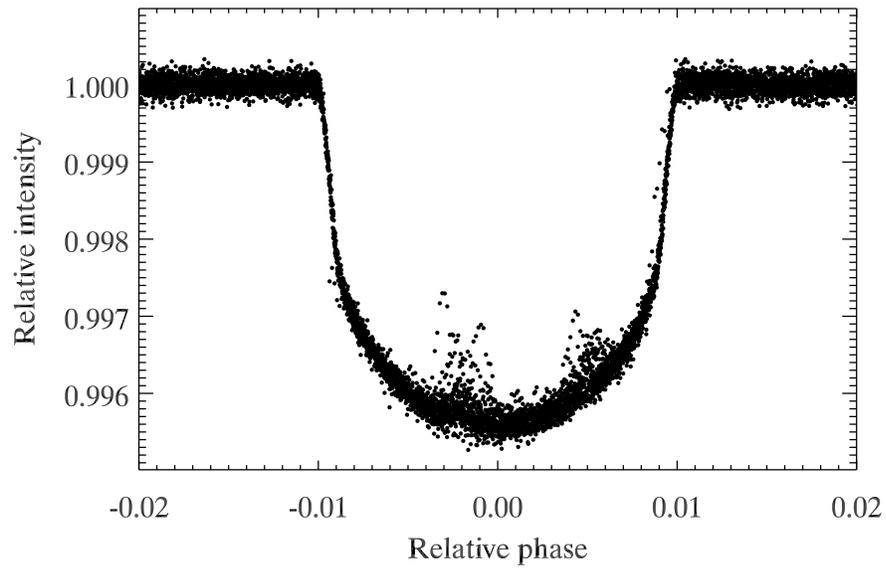}
% \vspace{0.8in}
\caption{Stack of 26 detrended and normalized transits of HAT-P-11
 from the Kepler Q0-Q2 data.  Note the tendency for perturbations by
 star spot crossings to concentrate at orbital phases near $-0.002$ \&
 $0.006$.  We attribute this to crossings of active latitudes by the
 planet in a highly inclined orbit \citep{winn, hirano}.
\label{fig4}}
\end{figure}

\clearpage

\begin{figure}
\epsscale{.70}
\plotone{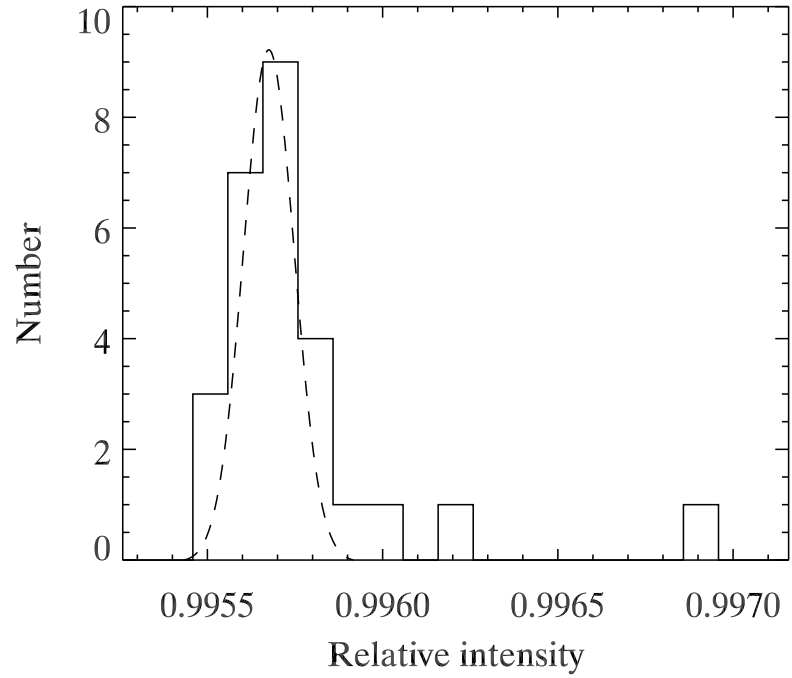}
% \vspace{0.8in}
\caption{Example of star spot removal at one particular phase in the
stacked HAT-P-11 transits of Figure~4.  The dashed line is a fit of a
Gaussian to the histogram of relative intensities at (in this example)
phase $-0.00098$.  The center of the Gaussian fit at intensity
$0.99567$ is the value of the transit curve at this phase with star spot crossings removed.
\label{fig5}}
\end{figure}

\clearpage

\begin{figure}
\epsscale{.70}
\plotone{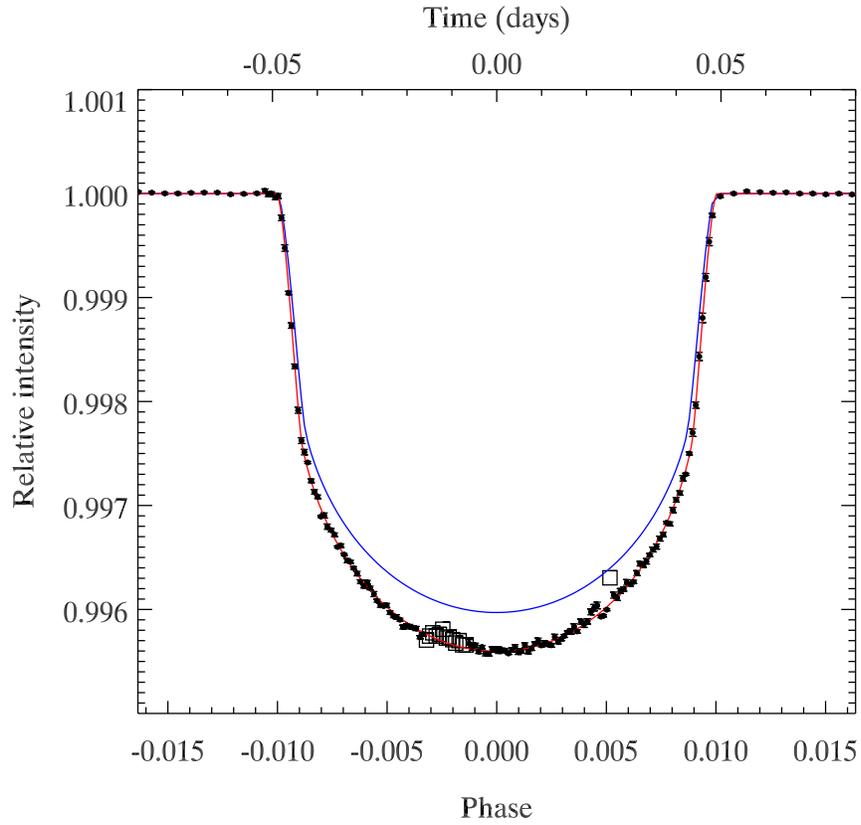}
% \vspace{0.8in}
\caption{Average Kepler transit curve for HAT-P-11, based on removing
star spot crossings and averaging the Figure~4 data.  The red line is
our best-fit MCMC solution (Kepler-1 line in Table~1). Error bars for
these Kepler data are plotted, but are difficult to discern, being
comparable to the size of the plot symbols.  The square points near
phase -0.0025 and the single point at +0.0055 were zero-weighted in
the fit. The blue curve is the transit expected using the B10 system
parameters, with the Kepler-1 limb darkening coefficients (Table~1).
\label{fig6}}
\end{figure}

\clearpage

\begin{figure}
\epsscale{.60}
\plotone{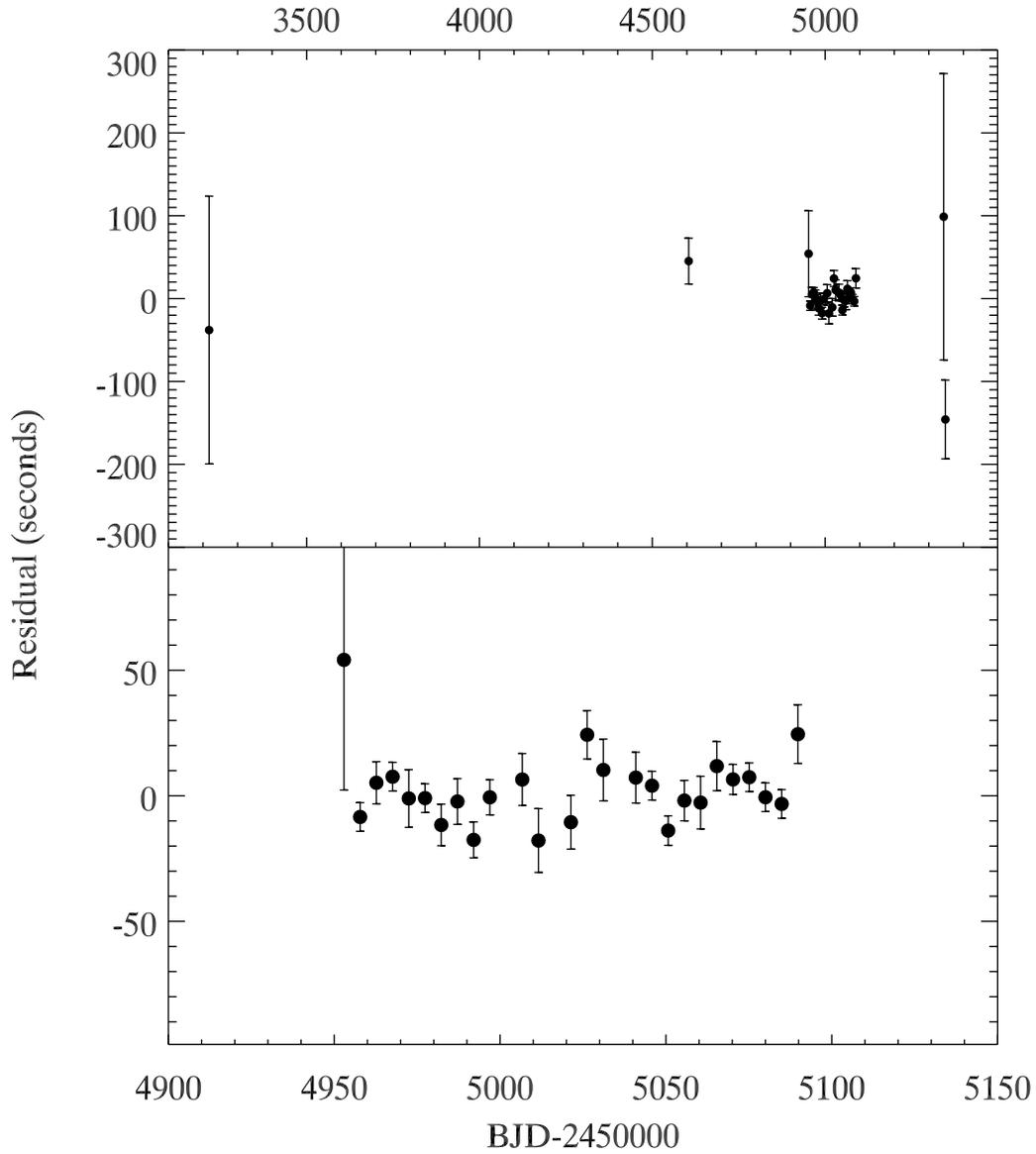}
\vspace{0.8in}
\caption{Transit time residuals for HAT-P-11 after removing the
best-fit ephemeris described in Sec.~6.  Top panel shows all data; the
two right-most points are from our ground-based data, and they are not
included in the ephemeris solution as noted in Sec.~2.1.  The cluster
of points near $BJD=2455000$ are the Kepler transits; they
dominate the ephemeris solution, and that time interval is expanded on the lower panel.
\label{fig7}}
\end{figure}

\clearpage

\begin{figure}
\epsscale{.7}
\plotone{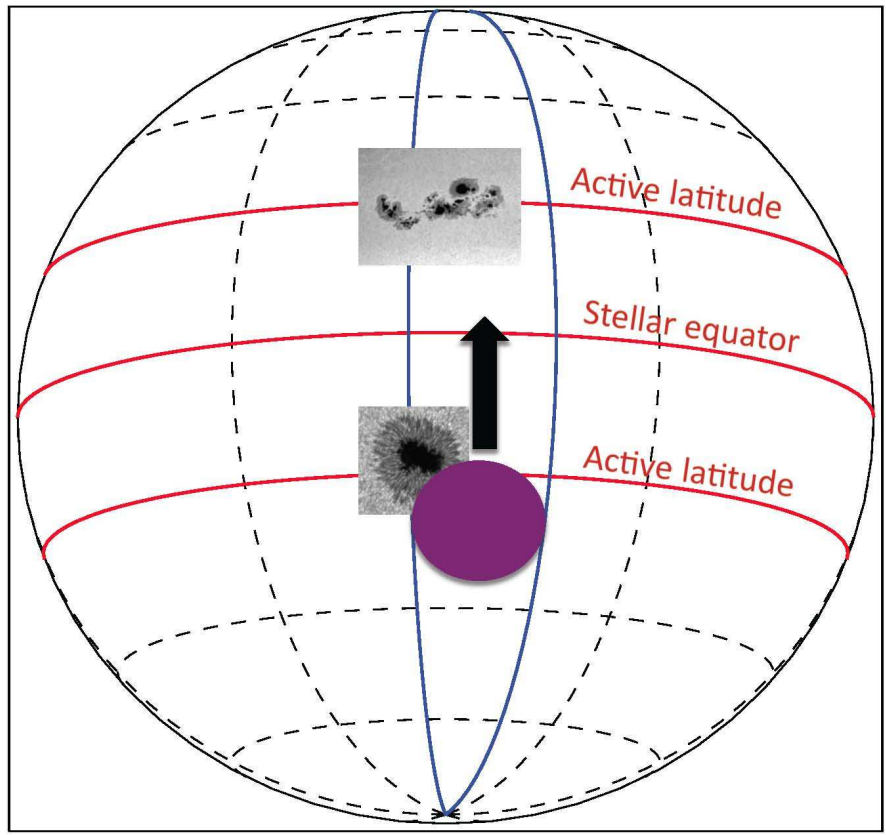}
\vspace{0.5in}
\caption{Cartoon of the HAT-P-11 transit geometry, illustrating that
the planet transits nearly perpendicular to the stellar equator, along
the sub-Earth longitude of the star.  The two blue meridians
illustrate that star spot regions crossed by the planet are
approximately bounded within a range of longitude (but not to exact
scale on this cartoon).  For clarity of the Sec.~7 discussion, the
path of the planet is here illustrated to be perpendicular to the
stellar equator.  While a perpendicular path is within the error limits of the
R-M results, we note that the sky-projected path is nominally tilted
by 103 degrees to the stellar equator \citep{winn}.
\label{fig8}}
\end{figure}

\clearpage

\begin{table}
\begin{center}
\caption{Results of MCMC fitting to HAT-P-11b transits. \label{tbl-1}}
\begin{tabular}{lllll}
\tableline
\tableline
Band and Methodology  & {\it u} & {\it i} & $R_p/R_s$ & $a/R_s$  \\
\tableline
Kepler-1  & $0.626$    & $89.41$    &  $0.05892$    & $16.549$  \\
  ---     & $\pm0.014$ & $\pm0.37$  & $\pm0.00027$  & $\pm0.230$ \\
         &             &             &               &            \\
Kepler-2  & $0.6179$    & $89.58$    &  $0.05865$    & $16.574$  \\
  ---     &  ---        & $\pm0.28$  & $\pm0.00017$  & $\pm0.156$ \\
          &             &             &               &            \\
J-band    & $0.086$     & $89.41$    &  $0.06274$     & $16.454$  \\
  ---     & $\pm0.065$  &  ---        & $\pm0.00066$  & $\pm0.131$ \\
          &             &             &               &             \\
B-band    & $1.000$     & $89.41$    &  $0.0596$    &  $16.549$  \\
  ---     & $\pm0.080$  &  ---        & $\pm0.0011$   &  ---   \\
\tableline

\end{tabular}
\end{center}
Note: Kepler-1 and -2 refer to different treatment of limb
darkening. The Kepler-1 line is our preferred solution; it fits to the linear
($u$) and quadratic limb darkening coefficients, whereas the Kepler-2
solution holds both $u$ and the quadratic coefficient fixed at
their Kurucz model atmosphere values (see text). Both the J-band and
B-band solutions hold the inclination fixed at the Kepler-1 value, and
the B-band solution also holds $a/R_s$ fixed at the Kepler-1 value.
\end{table}

\begin{table}
\begin{center}
\caption{HAT-P-11b transit times ($T_c$) from Kepler data. \label{tbl-2}}
\begin{tabular}{lll}
\tableline
\tableline
N  & $T_c$ (BJD-2450000)  & Error ($1\sigma$) \\
\tableline
72 &  4957.81318 &   0.00007 \\
73 &  4962.70114 &   0.00010 \\
74 &  4967.58897 &   0.00007 \\
75 &  4972.47667 &   0.00013 \\
76 &  4977.36448 &   0.00007  \\
77 &  4982.25215 &   0.00010 \\
78 &  4987.14006 &   0.00011 \\
79 &  4992.02769 &   0.00008 \\
80 &  4996.91569 &   0.00008 \\
82 &  5006.69137 &   0.00012 \\
83 &  5011.57889 &   0.00015 \\
85 &  5021.35458 &   0.00012 \\
86 &  5026.24279 &   0.00011 \\
87 &  5031.13042 &   0.00014 \\
89 &  5040.90599 &   0.00012 \\
90 &  5045.79376 &   0.00007 \\
91 &  5050.68135 &   0.00007 \\
92 &  5055.56929 &   0.00009 \\
93 &  5060.45709 &   0.00012 \\
94 &  5065.34506 &   0.00011 \\
95 &  5070.23280 &   0.00007 \\
96 &  5075.12061 &   0.00007 \\
97 &  5080.00832 &   0.00007 \\
98 &  5084.89609 &   0.00007 \\
99 &  5089.78421 &   0.00014 \\
\tableline
\end{tabular}
\end{center}
Note: Times are barycentric TDB \citep{eastman}, and our best-fit ephemeris, with
the ground based transits included, is $T_c=2454605.89155+4.8878018N$, as given in Sec.~6. 
\end{table}

%% If the table is more than one page long, the width of the table can vary
%% from page to page when the default \tablewidth is used, as below.  The
%% individual table widths for each page will be written to the log file; a
%% maximum tablewidth for the table can be computed from these values.
%% The \tablewidth argument can then be reset and the file reprocessed, so
%% that the table is of uniform width throughout. Try getting the widths
%% from the log file and changing the \tablewidth parameter to see how
%% adjusting this value affects table formatting.

%% The \dataset{} macro has also been applied to a few of the objects to
%% show how many observations can be tagged in a table.

\clearpage

%% Tables may also be prepared as separate files. See the accompanying
%% sample file table.tex for an example of an external table file.
%% To include an external file in your main document, use the \input
%% command. Uncomment the line below to include table.tex in this
%% sample file. (Note that you will need to comment out the \documentclass,
%% \begin{document}, and \end{document} commands from table.tex if you want
%% to include it in this document.)


%% The following command ends your manuscript. LaTeX will ignore any text

\end{document}